\documentclass[conference,a4paper,10pt]{IEEEtran}
\usepackage{cite}
\usepackage{amsfonts}
\usepackage{amsmath}
\usepackage[utf8]{inputenc}
\usepackage{hyperref}
\usepackage{amssymb,array}
\usepackage[english]{babel}
\usepackage[pdftex]{graphicx}

\usepackage{url}
\begin{document}

\title{Is the corporate elite disintegrating?\\
Interlock boards and the Mizruchi hypothesis}

\author{\IEEEauthorblockN{Kevin Mentzer}
\IEEEauthorblockA{\small Dept. of Mathematical Sciences and Dept. of CIS \\
Bentley University and Bryant University\\
Waltham, MA and Smithfield, RI, USA\\
Email: \url{kmentzer@bryant.edu }}
\and
\IEEEauthorblockN{Francois-Xavier Dudouet}
\IEEEauthorblockA{Dept. of Sociology\\
Université Dauphine\\
Paris, France\\
Email: \url{francois_xavier.dudouet@dauphine.fr}}
\and
\IEEEauthorblockN{Dominique Haughton}
\IEEEauthorblockA{\small Dept. of Mathematical Sciences, Bentley University \\
Waltham, MA, USA\\
SAMM, Université Paris 1, Paris, France\\
GREMAQ, Université Toulouse 1, Toulouse, France \\
Email: \url{dhaughton@bentley.edu} }
\and
\IEEEauthorblockN{Pierre Latouche}
\IEEEauthorblockA{\small SAMM \\
Universit\'e Paris 1 Pantheon-Sorbonne\\
Paris, France 75634 Paris Cedex 13\\
Email: \url{pierre.latouche@univ-paris1.fr}}
\and
\IEEEauthorblockN{Fabrice Rossi}
\IEEEauthorblockA{\small SAMM \\
Universit\'e Paris 1 Pantheon-Sorbonne\\
Paris, France 75634 Paris Cedex 13\\
Email: \url{Fabrice.Rossi@univ-paris1.fr}}
}

\maketitle

\begin{abstract}
  This paper proposes an approach for comparing interlocked board networks
  over time to test for statistically significant change. In addition to
  contributing to the conversation about whether the Mizruchi hypothesis (that
  a disintegration of power is occurring within the corporate elite) holds or
  not, we propose novel methods to handle a longitudinal investigation of a
  series of social networks where the nodes undergo a few modifications at
  each time point. Methodologically, our contribution is two-fold: we extend a
  Bayesian model hereto applied to compare two time periods to a longer time
  period, and we define and employ the concept of a hull of a sequence of
  social networks, which makes it possible to circumvent the problem of
  changing nodes over time.
\end{abstract}

\IEEEpeerreviewmaketitle

\section{Introduction}
In The Fracturing of the American Corporate Elite \cite{mizruchi13:_americ}, Mizruchi makes the
claim that a disintegration of power is occurring within the corporate
elite. He asserts that since World War II the voice of the American corporate
elite has diminished to the point where it is now ineffectual. This shift in
the role of the corporate elite has had an impact not only on the corporations
it oversees but on the entire business system \cite{scott91:_networ,useem84}.

We argue that if this is occurring then there should be evidence of such
change in the power networks of members of the corporate elite. One way in
which to identify the corporate elite is to look at corporate boards of
directors. Building on the work of Mills \cite{mills56}, extensive research about the
power elite has been conducted, including investigations of the influence of
board members on issues related to executive compensation and firm
performance.  Wong et al. \cite{Wong201585} show that the level of board interlocking
contributes to the level of executive pay. 

In this paper, we focus on interlocked boards - boards that share at least one
director. A premise to our study is to consider the density of an interlock
network as a proxy for cohesive power within the corporate elite. 

Employing a Bayesian analysis in which we test for significance in network
change over time, we show that there is little evidence that degradation in
the cohesiveness of the network is occurring within the elite of the elite. We
use the Dow Jones 30 (DOW 30) group of corporations to represent the ``elite of
the elite'' and our period of study is 2001-2010 (Section III). This result
suggests that the Dow Jones 30 forms a very robust core which is rather
impervious to degradation from outside forces.    

We thus demonstrate how to apply a Bayesian model to better understand changes
over time in an interlocked corporate board network. This work builds on the
Bayesian version of the $p_1$ model introduced by Wong \cite{wong87:_bayes_model_direc_graph}, extended by Gill and
Swartz \cite{paramjit04:_bayes_analy_direc_graph_data}, and applied by Adams et al. \cite{adams08:_chang_connec_social_networ_time} to changes in the density of
collaborative networks over two periods of time. This extended p1 model
includes random effects allowing for some network internal dependency (see
also Goldenberg et al. \cite{goldenberg2010survey} for a useful review paper).  

In this paper, we report on successive pairwise comparisons of the networks
(Section IV) and then propose in Section V a method to perform a full
longitudinal analysis.

\section{Literature and background }
In The Power Elite, C. Wright Mills \cite{mills56} discusses the structural changes
occurring within the United States in the 1930’s and 1940’s that gave rise to
the power elite. He identifies three changes that make this rise in power
possible: 1. the increasing dominance of corporations, 2. the expansion of the
federal government, and 3. the emergence of a large military body following
World War II. However, Mizruchi, in many articles but most visibly in \cite{mizruchi13:_americ},
argues that while the corporate elite was prominent in helping to direct the
public agenda through World War II, its voice has since that time become
fragmented and has lost its power.  

Researchers often use interlocked board membership as the identifying
characteristic to represent the corporate elite. It has been argued that
members of interlocked boards are able to move in and out of not only board
rooms but also of civic arenas to spread their beliefs. Their board prominence
also gives them the recognition to be seen as senior statesmen in their
communities. 

This “inner circle” thus possesses a higher level of political influence and
social cohesion \cite{useem84}. This implication has led to the realization that
corporations do not exist in isolation, but instead are part of a societal
power establishment through individuals including interlocked board members
\cite{carroll14:_sage_handb_social_networ_analy}.

Another interesting question is of whether a core corporate group such as the
Dow Jones constitutes a ``rich club'' in the sense described for example in
\cite{alstott14:_unify_framew_measur_weigh_rich_clubs}. This would imply that corporations within the Dow Jones interlock with
each other more than with corporations outside the Dow Jones.

The similarities of powers obtained by the corporate elite are not the only
similarities. Others \cite{stanworth75:_the_moder_corpor_econom,whitley74} have shown that demographic similarities can be
observed within these members as well. They found that, when considering
education and social characteristics, interlocked members have become more
similar over time. One might therefore expect that these elite members will
express similar opinions regarding societal directions, in turn helping to
contribute to the singular voice of the elite.   

In the United States, the most important initiative undertaken to limit
interlocks is the Clayton Act of 1914 which prohibits U.S. firms that compete
with one another from sharing board members. 

Board member turnover is low, with the average board member tenure being
approximately 12 years \cite{vafeas03:_lengt_board_tenur_outsid_direc_indep}. Because of this we don’t expect to see
significant change in the interlocked network on a year-by-year
basis. However, we would not be surprised if events related to the Clayton
Act, as described above, could result in periods of disruption in the
interlock network.

In this paper, we investigate interlock networks under the lens of Social
Network Analysis (SNA). A link occurs between corporations when they share a
board member, or between directors where they serve on the same board. Such a
network is often referred to as a bipartite network. 

\section{Data and proposed model}
\subsection{Data Preparation}
ver the 128-year history  of the Dow Jones Industrial Average the companies
that make up the index have changed 53 times. The index started in 1884 with
12 companies. This list expanded to 20 companies in 1916 and expanded again in
1928 to include a total of 30 companies, where it stands today. Our period of
interest is from 2001 through 2010 which included 43 total companies, due to
additions and subtractions over this period. We will return in Section V to a
discussion of this group, which we will refer to as the hull of the DOW 30.

Having identified these 43 companies, our next step was to filter the list of
directors down to just those directors who served on the boards of these
companies. This allowed us to condense all of the directors of interest into a
single file. 

The data were obtained from the BoardEX database, which provides a row for
each director who overlapped with another director in the same firm for some
period of time (the overlap period). A new row would be created if either
director changed positions within the firm, or changed director roles. The
overlap period was difficult to work with, and had to be segmented into a
start and end date. Eventually we were able to get the file down to a unique
list of directors for each firm during only the period of interest. 

\begin{figure}[htbp]
  \centering
\includegraphics[width=\linewidth]{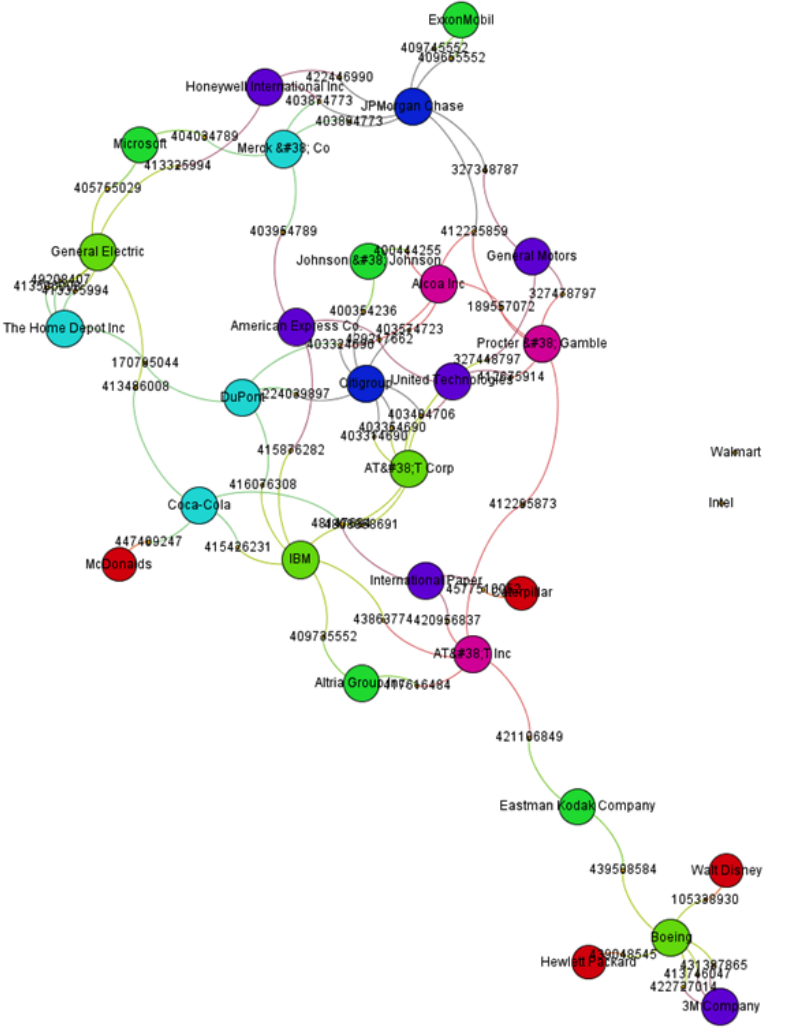}  
  \caption{Interlocked network showing interlocked board members, 2001}
  \label{fig:interlock:2001}
\end{figure}

From this list we were able to create our nodes and edges. We started by
creating nodes for both the companies as well as the directors and creating
the edges between them. This allowed us to see demographics on specific
directors (age, gender, tenure, role, etc.) which opened the way for us to
manually spot check our data using director lists found in the annual reports
of these companies. We had to rely on these demographics instead of board
member names due to the lack of names in the BoardEX database. Our network, at
this stage, can be seen for 2001 in Figure \ref{fig:interlock:2001}, with the nodes sized on the
basis on their degree centrality (note that we have displayed all 30 companies
regardless of whether they had a shared board member, but we have filtered out
board members that were not considered interlocked to keep the graph as easy
to read as possible). We recall that the graph in Figure \ref{fig:interlock:2001} is a bipartite, or
2-mode, graph since it contains two types of nodes, companies and directors.

\begin{figure}[htbp]
  \centering
 \includegraphics[width=\linewidth]{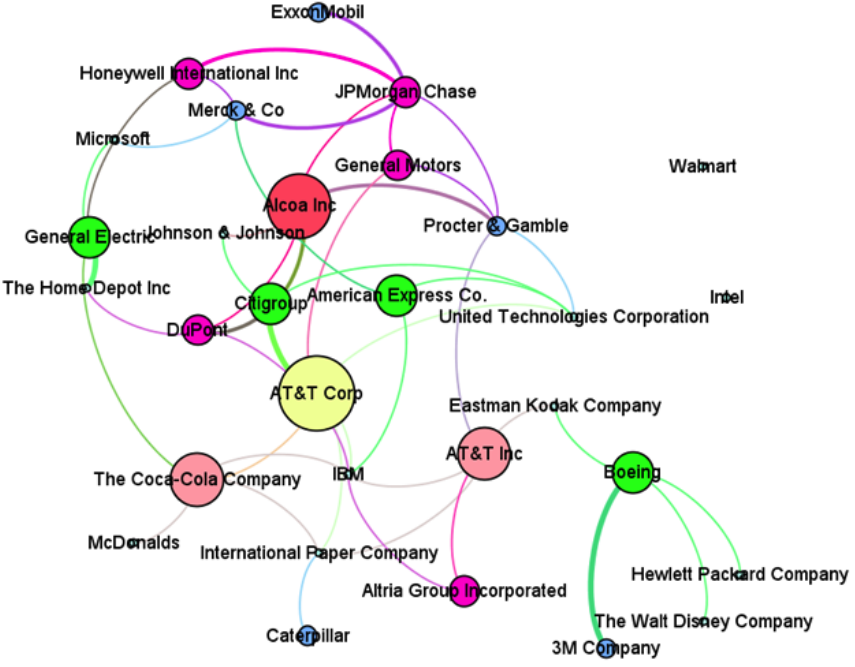}   
  \caption{DOW 30 interlocked boards’ network 2001; colors of nodes represent
    weighted degree}
  \label{fig:dow30:interlock}
\end{figure}

Next, in Figure \ref{fig:dow30:interlock}, we have stripped out the individual board members and have
replaced them with weighted connections between the two companies, which
converts our 2-mode network into a 1-mode network. So, as can be seen in
Figure \ref{fig:dow30:interlock}, General Electric (GE) and Microsoft have one common board member so
an edge with a weight of 1 was created between GE and Microsoft, while GE and
Home Depot share two board members which resulted in an edge with a weight of
2 between GE and Home Depot. The 2001 network for the DOW 30 companies can be
seen in Figure \ref{fig:dow30:interlock}. 

\subsection{Proposed model}
Following a model suggested by Wong \cite{wong87:_bayes_model_direc_graph} and Gill and Swartz \cite{paramjit04:_bayes_analy_direc_graph_data}, we employ a
Bayesian methodology in a search for statistical significance in the changes
in density of social networks over time. Testing for statistical significance
over time has been of interest to the social network research community due to
the complexities faced, including link dependency. In addition to link
dependency, snapshots of the network over time are typically also dependent
and, therefore, traditional means that require independent observations cannot
be employed. The model can be described as follows: $Y^k_{ij1}=1$ and
$Y^k_{ij0}=0$, if $i$ and $j$ are connected, and $Y^k_{ij1}=0$ and $Y^k_{ij0}=1$,
if $i$ and $j$ are not connected, with
\begin{equation*}
\ln P(Y^k_{ij0}=1)=\lambda^k_{ij},
\end{equation*}
and
\begin{equation*}
\ln P(Y^k_{ij1}=1)=\lambda^k_{ij}+\theta^k+\alpha_i^k+\alpha_j^k.
\end{equation*}
The index $k$ denotes the time period and indices $i$ and $j$ refer to two
corporations.  Because each pair of corporations $(i, j)$ can either share or
not share a board member, the matrix $Y^k_{ij0}$ is a simple opposite of the
matrix $Y^k_{ij1}$ in the sense that $Y^k_{ij0}$ can be obtained from
$Y^k_{ij1}$ by replacing zeros with ones and ones with zeros. The matrix
$Y^k_{ij1}$ is often referred to as the sociomatrix, with its ones indicating
where a link occurs. The probability $P(Y^k_{ij1}=1)$ represents the
probability of an interlock link occurring between corporations $i$ and $j$,
at time $k$ ,and $P(Y^k_{ij0}=1)$ represents the probability that no such link
exists. The parameter $\theta^k$ represents the overall propensity for links
to occur in the network at time $k$, and the parameters $\alpha^k_i$ represent the
propensity for corporation $i$ to share board members with other corporations
in the network. Prior distributions are defined on each parameter, making the
model Bayesian.

The advantage of this approach to modelling links in a sequence of networks is
that it allows for links to not be independent (via the random effects
$\alpha_i^k$ ) at a given point in time and also allows, via suitable priors,
for parameters to exhibit a timewise correlation: for example, it is quite
likely that the $\theta^k$ and the $\alpha_i^k$ are correlated over time. We
will return to this point in Section V.

In Adams et al. \cite{adams08:_chang_connec_social_networ_time} this approach was used successfully to examine changes in
network parameters over two time periods. In this paper we explain how to
extend this approach to more than two time periods, with a more complicated
time correlation structure, such as autoregressive, for example. We have
implemented the Bayesian model for the two time periods 2001 and 2002, and for
each successive pair-wise comparison (2002-2003, 2003-2004, etc.) as well as
2001-2010 (Section IV) and conducted a preliminary time series analysis in
Section V. 

Our DOW 30 components change over time, which presents a unique challenge to
modeling the network. We have several years of stability where the same 30
companies are present, but in other years some companies are replaced with
others. Over the 10 year period under study the following changes occurred: 

\begin{enumerate}
\item In 2004 AT\&T Corporation, Eastman Kodak, and International Paper
  were all removed while American International Group Inc, Pfizer, and Verizon
  were added.
\item  In 2008 Altria, American International Group, and Honeywell were replaced
 with Bank of America, Chevron, and Kraft Foods.
\item In 2009 Citigroup and General Motors were replaced with Cisco Systems
  and Travelers.
\end{enumerate}
It follows that we can compare years within each group 2001-2003, 2004-2007,
and 2009-2010, but comparisons across these time periods may present a
challenge.

\begin{table*}[t]
  \centering
\begin{tabular}{lllllllllll}
&2001&2002&2003&2004&2005&2006&2007&2008&2009&2010\\\hline
\# Nodes&30&30&30&30&30&30&30&30&30&30\\
\# Edges&46&46&48&51&47&46&48&44&40&50\\
Avg Degree&3.067&3.067&3.2&3.4&3.133&3.067&3.2&2.933&2.667&3.333\\
Avg Weighted Degree&3.933&3.867&3.867&4.067&3.533&3.533&3.467&3.2&2.8&3.667\\
Network Diameter&7&7&5&5&5&6&5&6&7&5\\
Graph Density&0.106&0.106&0.11&0.117&0.108&0.106&0.11&0.101&0.092&0.115\\
Modularity&0.497&0.478&0.437&0.441&0.402&0.454&0.433&0.529&0.465&0.447\\
Connected Components&3&3&4&3&3&2&2&1&3&3\\
Avg Clustering Coefficient&0.226&0.169&0.324&0.336&0.168&0.212&0.197&0.292&0.263&0.235\\
Avg Path Length&3.071&2.934&2.584&2.495&2.54&2.845&2.645&3.062&3.114&2.514\\\hline
\end{tabular}
  \caption{Key network level measures dow-30, 2001-2010}
  \label{tab:keyvalue}
\end{table*}

\begin{figure}[htbp]
  \centering
\includegraphics[width=\linewidth]{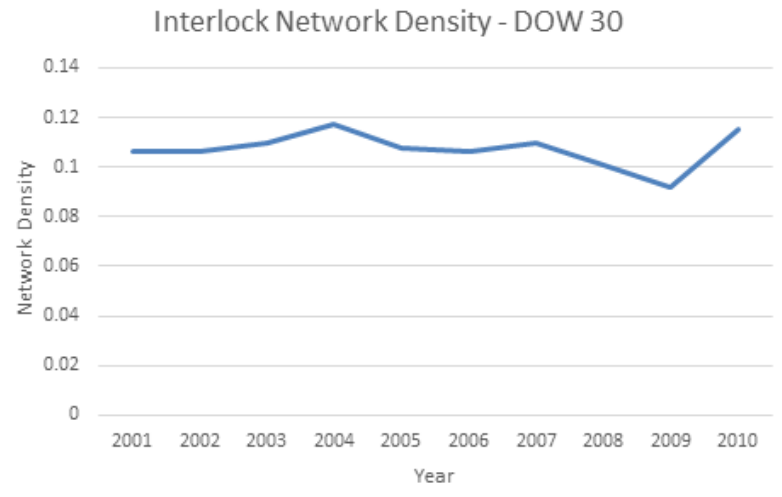}    
  \caption{Interlock network density, 2001-2010}
  \label{fig:interlock:density}
\end{figure}

\section{Analysis and results}
Table \ref{tab:keyvalue} provides key network level measures for 2001-2010 and Figure \ref{fig:interlock:density}
graphically displays the evolution of the density of the networks.  It is
clear that observed network measures at first sight remain essentially stable
over the period of interest. One might note a slight disintegration effect in
2008 and 2009, near the time of highest intensity of the recent financial
crisis.  

\begin{figure}[htbp]
  \centering
\includegraphics[width=\linewidth]{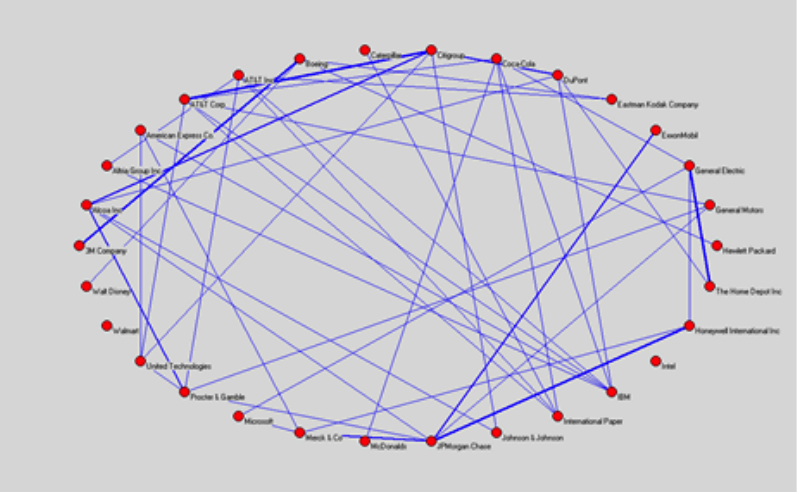}      
  \caption{DOW 30 interlock network 2001}
  \label{fig:dow30:2001}
\end{figure}

\begin{figure}[htbp]
  \centering
\includegraphics[width=\linewidth]{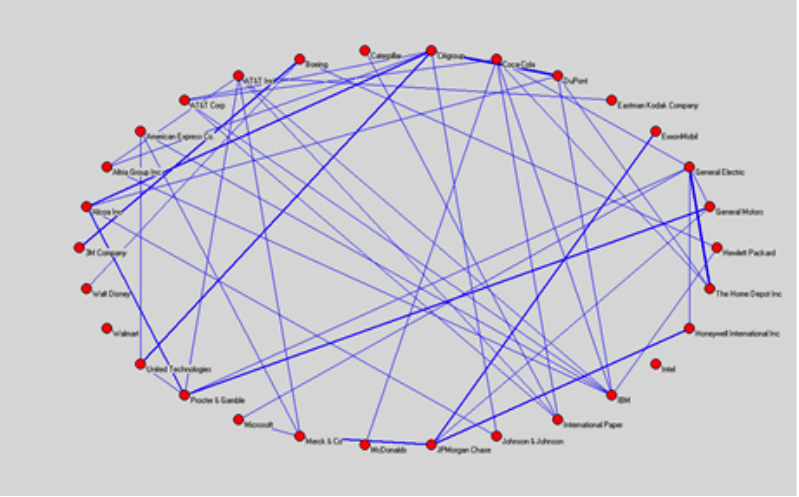}      
  \caption{DOW 30 interlock network 2002}
  \label{fig:dow30:2002}
\end{figure}

Focusing, for example, on the pair comparison between 2001 and 2002, we can
visually observe few differences between the two network graphs for 2001 and
2002 in Figures \ref{fig:dow30:2001} and \ref{fig:dow30:2002}. 

We next employ our Bayesian model to investigate whether our visual
observations are confirmed by an examination of the posterior distribution of
the difference in cohesiveness between 2001 and 2002, and then between all
other successive pairs of years, ending in 2009-2010.  

\begin{figure}[htbp]
  \centering
\includegraphics[width=0.8\linewidth]{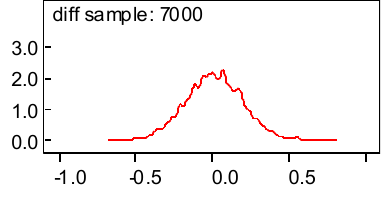}  

\includegraphics[width=0.8\linewidth]{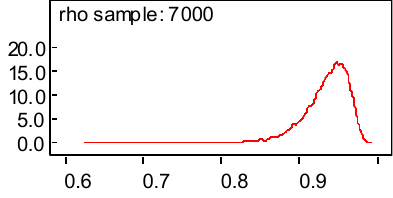}  
  \caption{Kernel posterior densities for the difference between 2001 and 2002
    in propensities to form links and for the correlation in network
    parameters between 2001 and 2002.}
  \label{fig:posterior:2001}
\end{figure}

Figure 6 displays graphs of the kernel posterior densities for the difference
in the propensity to form links (cohesiveness, playing here is the role of
density) in the networks for 2001 and 2002, and for the correlation of the
network parameters between 2001 and 2002. Since the value of zero is in the
middle of the posterior distribution for the difference, that would imply no
significant change between the cohesiveness of the 2001 and 2002 networks.  It
also appears, as expected, that the correlation is high, with a posterior
distribution centered tightly above .9. 

The posterior mean of the difference between 2001 and 2002 is .01548, with a
2.5\% - 97.5\% credible interval of (-0.3584, 0.3923) encompassing zero. The
positive value of the posterior mean of the difference indicates a decrease in
the cohesiveness of the network between 2001 and 2002, but this difference is
a posteriori essentially as likely to be positive as to be negative.  A more
pronounced difference would have its posterior density shifted away from zero.

\begin{figure}[htbp]
  \centering
\includegraphics[width=0.8\linewidth]{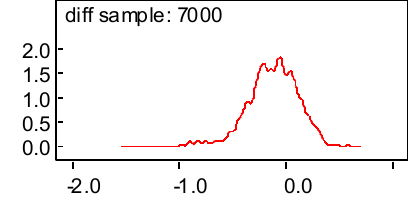}  \includegraphics[width=0.8\linewidth]{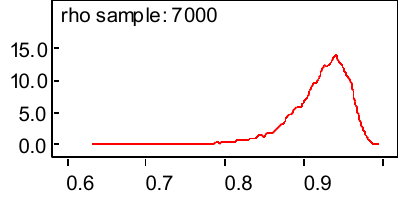}  
  \caption{Kernel posterior densities for the difference between 2002 and 2003
    in propensities to form links and for the correlation in network
    parameters between 2001 and 2002.} 
  \label{fig:posterior:2002}
\end{figure}

The posterior mean of the difference between 2002 and 2003 of -.1224 indicates
an increase in the cohesiveness of the network between 2001 and 2002, but with
a 2.5\% - 97.5\% credible interval of (-0.7454, 0.3104) again encompassing
zero.

A similar situation emerges among the remaining pairwise comparisons,
including the comparison of 2001 with 2010 (beginning and end of studied
period). The posterior densities vary a bit in how central zero is to the
distribution, but overall, one does not detect any changes that might be
referred to as significant in a statistical sense. 

On the basis of our analysis to date we see no trend away from interlocked
networks over the time period studied. So while we do not find support for
Mizruchi’s claim, we feel the time period needs to be expanded to get a true
sense of the network changes over decades.

\section{Extended dow jones and longitudinal interlock network}
As discussed earlier, one complication in longitudinal investigations of the
DOW 30 network is that corporations leave and enter this network as years
pass.  

We therefore propose to employ a DOW 30 hull which includes all corporations
that have appeared at least once in DOW 30 during the period of study (43
corporations for the period 2001-2010). We construct a network on this hull
consisting of all links connecting corporations in the hull who at any stage
shared a director. All directors, serving at any stage on boards of
corporations featured in the hull, are nodes in the bipartite network.  We
suggest that this is an appropriate way to identify a corporate “elite of
elite” which is somewhat more robust than the DOW 30 strictly defined.  

This makes it possible to build longitudinal models of networks with a fixed
set of nodes, and to compare and extend existing techniques.  As a first foray
into the temporal behavior of the series of networks, we extract the estimated
differences (posterior means) from each of the pairs 2001-2002, 2003-2004, …
2009-2010 and fit a simple AR(2) (Auto-Regressive of order 2) model to the
time series of the 9 differences. The fit of this preliminarily model is
displayed in Figure 8. It is interesting to note that observation 8 is the
pair 2008-2009 when the financial crisis was probably at its most severe
level.

\begin{figure}[htbp]
  \centering
\includegraphics{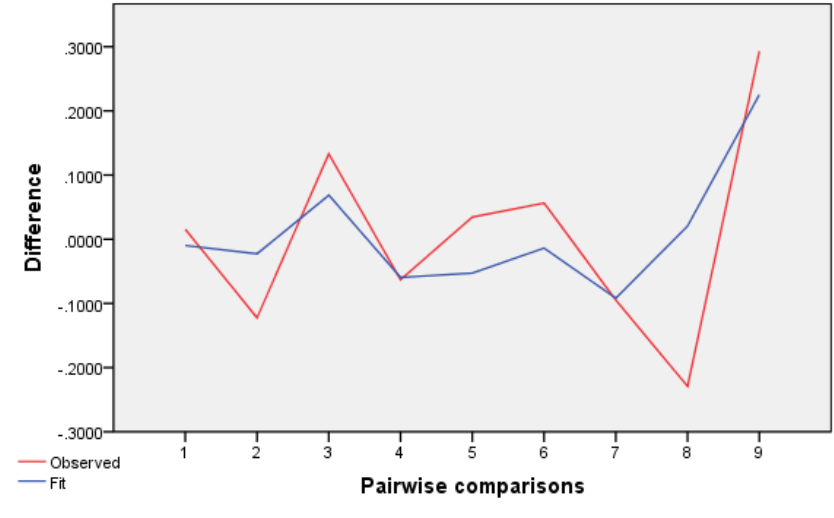}  
  \caption{AR(2) model on the nine estimated pair differences}
  \label{fig:AR}
\end{figure}

The AR(2) model does a fairly good job of explaining correlations across the
years (the root mean square error RMSE is of .115 and BIC of -3.59 for the
AR(2) model, compared to a RMSE of .135 and a BIC of -3.51 for an AR(1)
model), but we believe that more sophisticated models constructed on the DOW
30 hull will reveal more aspects of the evolution of the core interlock
network over time. We are also aware that our current study period is limited,
and expect to extend this period to a much longer time span; from World War II
to the present.

Using the hull of the sequence of DOW 30 interlock networks, we also expect to
be able to take into account in our model the weights of the links (related to
the number of shared board members).

\section{In conclusion}
This paper has proposed an approach that can be employed to compare
interlocked board networks over time to test for statistical significant
change. Our preliminary findings suggest that change within the network does
not usually occur rapidly enough to be detected in a year-over-year
comparison. This is not surprising given the size of the network (number of
nodes) and the frequency with which directors for any company change in a year
over year basis. This does not mean that no shocks have caused changes over
the years, but we have just not yet detected them in our analysis to date.

While we were not surprised to not find significant change on a year to year
basis, we were surprised that even our 10 year analysis (2001 versus 2010)
also displayed no significant change. This suggests that the entire decade was
stable. Since Mizruchi claims that this fracturing has been occurring since
World War II, our next step is to expand the analysis to go back decades in
order to see when this change may have occurred.

In addition to contributing to the conversation on whether the Mizruchi
hypothesis holds or not, we have proposed novel methods to handle a
longitudinal investigation of a sequence of social networks where the nodes
undergo a few modifications at each time point. Methodologically, our
contribution is two-fold: we extend a Bayesian model hereto applied to compare
two time periods to a longer time period, and we define and employ the concept
of a hull of a sequence of social networks, which makes it possible to
circumvent the problem of changing nodes over time.

\bibliographystyle{IEEEtran}

\bibliography{IEEEabrv,mybibfile}

\end{document}